\documentclass[twocolumn]{aastex62}
\usepackage[utf8]{inputenc}
\usepackage{hyperref,astrojournals,amsmath,amssymb,mathtools,graphicx,txfonts,microtype,nicefrac,xspace,bm}
\usepackage{enumitem,kantlipsum}

\hypersetup{%
  pdftitle={Apsidal Alignment},
  pdfauthor={\textcopyright\ authors},
  bookmarksopen=true,
  colorlinks=true,
  linkcolor=black,
  citecolor=black,
  urlcolor=black}

\newcommand{\Md}{M_{\rm disk}}

\renewcommand{\c}[2]{\ensuremath{#1_{\text{#2}}}} 
\newcommand{\ia}{\c{i}{a}\xspace}
\newcommand{\ib}{\c{i}{b}\xspace}

\begin{document}

\title{Steeper Scattered Disks Buckle Faster}

\shortauthors{Zderic \& Madigan}

\author[0000-0003-2961-4009]{Alexander Zderic}
\affiliation{JILA and Department of Astrophysical and Planetary Sciences, CU Boulder, Boulder, CO 80309, USA}
\email{alexander.zderic@colorado.edu}

\author[0000-0002-1119-5769]{Ann-Marie Madigan}
\affiliation{JILA and Department of Astrophysical and Planetary Sciences, CU Boulder, Boulder, CO 80309, USA}

\begin{abstract}

Disks of low-mass bodies scattered by giant planets to large semi-major axis and constant periapsis orbits are vulnerable to a buckling instability. This  instability exponentially grows orbital inclinations, raises periapsis distances, and coherently tilts orbits resulting in clustering of arguments of periapsis. The dynamically hot system is then susceptible to the formation of a lopsided mode. 
Here we show 
that the timescale of the buckling instability decreases as the radial surface density of the population becomes more centrally dense, i.e., steeper scattered disks buckle faster.
Accounting for differential apsidal precession driven by giant planets, we find that $\sim\!10\,M_\oplus$ is sufficient for a primordial scattered disk in the trans-Neptunian region to have been unstable if $dN \propto a^{-2.5} da$.
    \newline
\end{abstract}
\section{Introduction}

Axisymmetric disks of bodies on eccentric orbits in a near-Keplerian potential are collectively unstable to an out-of-plane buckling \citep{Madigan2016}. The dynamics are driven by orbit-averaged torques between the massive disk bodies and result in exponential growth of orbit inclinations, a decrease in orbital eccentricities, and clustering in arguments of periapsis \citep{Madigan2018}. This `inclination instability' results in a hotter system which is susceptible to the formation of a lopsided mode or clustering in longitude of periapsis  \citep{Zderic2020a,Zderic2021}. 

We have suggested that this collective instability could explain the orbital anomalies observed in the trans-Neptunian outer solar system which include detached objects, extreme inclination orbits, and possible clustering in arguments of periapsis and longitude of periapsis
\citep{Trujillo2014, Batygin2016,Shankman2017,Lawler2017,Becker2018,Brown2019, Bernardinelli2020,Kavelaars2020,Napier2021}. Migrating giant planets interacting with planetesimals produce massive scattered disks early in the solar system's evolution \citep{Duncan1997,Vokrouhlicky2019}. These eccentric primordial scattered disks are susceptible to the inclination instability through mutual orbit-averaged torques.

The gravitational torques that drive the instability are sensitive to external perturbations. The differential precession induced in the trans-Neptunian region by the outer giant planets reduces the strength of inter-orbit torques and lengthens the instability timescale. \citet{Zderic2020b} found that the total disk mass between $100 - 1000$ au must have been $\gtrsim 20\,M_\oplus$ at some time in the past for the instability to occur. This was based on $N$-body simulations of a primordial scattered disk with a relatively shallow surface density profile, $dN \sim a^{-1} da$. 
Here we show that the steepness of the radial mass distribution strongly affects the timescale of the inclination instability, with more centrally dense mass distributions having significantly faster timescales. 
We run a suite of $N$-body simulations of disks undergoing the inclination instability with a range of radial mass distributions, and find that the timescale of the inclination instability varies with the median period within the disk. There is a factor of $\gtrsim \! 6$ difference in timescales across the range of power law indices explored.
In \citet{Zderic2020b}, we found that timescale of the inclination instability plays an important role in determining the instability's robustness to external sources of differential orbital precession.
Here we find that disks with steeper radial distributions require less mass to be unstable to the inclination instability.
For example, \citet{Huang2022} generate a primordial scattered disk using simple Neptune-scattering simulations with test particles finding $dN \sim a^{-2.5} da$ after $50\,{\rm Myr}$.  
With this steeper distribution, $\sim\!10\,M_\oplus$ is sufficient for the inclination instability to have occurred in the primordial trans-Neptunian region.

\section{Simulations and Timescales}
\label{sec:simulations}

Here we run $N$-body simulations of unstable scattered disks to quantify the effect of varying the radial mass profile.  
Our simulations are run using the {\tt REBOUND} $N$-body integrator framework \citep{Rein2012, Tamayo2020}.
We use the IAS15 integrator with interactive, massive particles. 
We run simple simulations to directly compare with our previous results. 
The $N$-body disks have $\Md = 10^{-3}\,M_\odot$ and $N = 400$ with each particle having identical mass, $m = \Md / N$. 
The unrealistically large disk mass accelerates the secular dynamics which reduces the computing time required to capture the instability, and the low particle number is required for these relatively inefficient but highly accurate simulations.\footnote{Fixed-timestep, hybrid-symplectic integrators fail to resolve the highly eccentric orbits in our simulations.  However, walltimes for $N\gtrsim10^3$ with the IAS15 integrator for our disks are prohibitively long.}
The artificially large $\Md$ and low $N$ means we cannot add the giant planets to the simulation as the back-reaction of the disk on the giant planets would be unphysically large.
We discuss how these results apply to the actual primordial trans-Neptunian population, including the influences of the giant planets, in Section~\ref{sec:discussion}.

The initial disk orbital parameters are as follows:
the periapsis distances are $30\,{\rm au}$, inclinations ($i$) are drawn from a Rayleigh distribution with a mean inclination of $5^\circ$, and the arguments of periapsis ($\omega$), longitudes of the ascending node ($\Omega$), and mean anomalies are drawn from a uniform distribution in $[0,360^\circ)$.
The semi-major axes of the orbits are drawn from a 1D power law, $dN \propto a^{-\alpha} da$, in the range $[a_i, 10a_i]$ where $a_i = 100\,{\rm au}$ (with constant periapsis, $e \in [0.7,0.97]$).
As the disks are composed of eccentric orbits, the semi-major axis distribution is different from the radial distribution. However, $a^{-\alpha} \approx r^{-\alpha}$ outside of the apoapsis of the innermost orbit. 
The simplicity of these simulations allow us to directly compare with our previous work and to apply these results at a later date to exoplanet systems of varying orbital architecture. 
Future work applying the instability to the outer solar system will need to incorporate scattered disks created by the four (or more) migrating gas and ice giants.  

\begin{figure}[!t]
    \centering
    \includegraphics[width=\columnwidth]{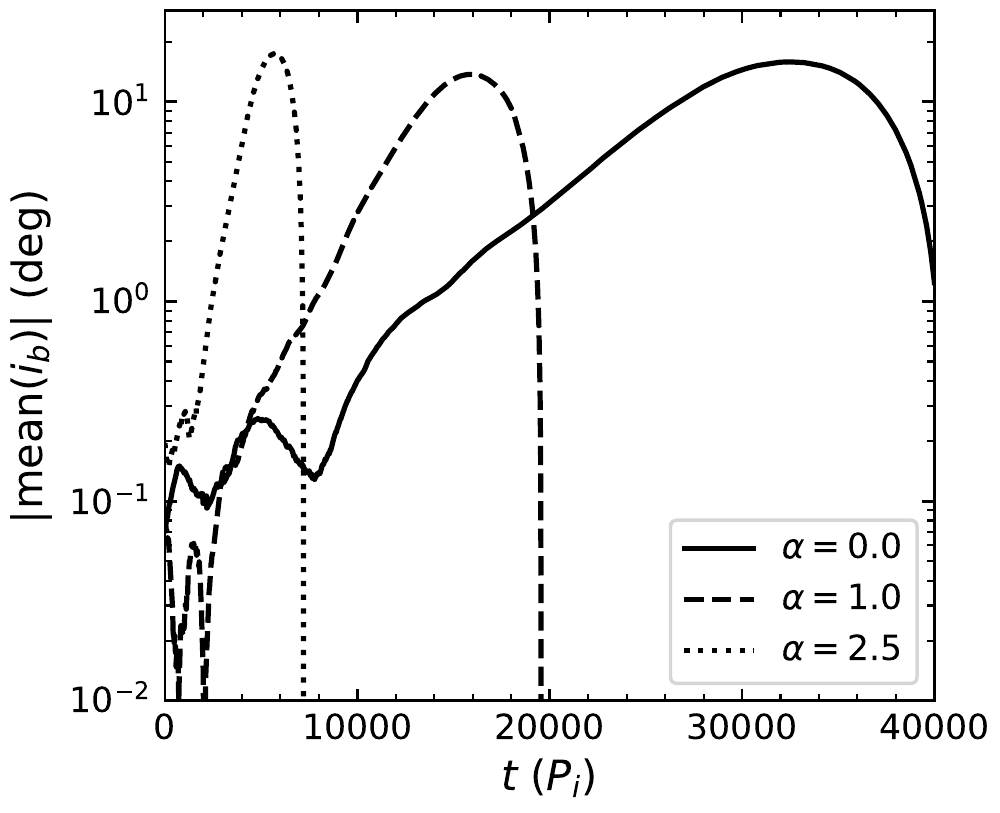}
    \caption{Absolute value of the mean $\ib$ for three  disks with different semi-major axis distribution indices, $\alpha$, and identical $N$ and $\Md$. We've truncated points after the linear stage of the instability to emphasize the growth rate differences. The timescale for the instability clearly decreases with increasing $\alpha$ (steeper density profiles).}
    \label{fig:growth-rate-compare}
\end{figure}

The collective dynamics behind the instability are described in \citet{Madigan2018}. In brief, we can understand the mechanism by considering just two eccentric orbits in a disk. A small vertical perturbing force will induce a torque on an eccentric orbit which causes it to roll over its major axis. This rolling in turn induces a torque on a nearby orbit causing it to pitch over its semi-latus rectum. For sufficiently eccentric orbits ($e \gtrsim 0.5$), the pitching acts to raise or lower the apocenter of the second orbit such that it reinforces the initial vertical force. One can solve linearized equations of motion of the system and demonstrate exponential growth of orbital inclinations in the initially thin disk. 
The coherent pitching and rolling of the orbits in the same direction means that the apocenters of the orbits all rise above or drop below the mid-plane together. 
The disk buckles into a cone or bowl shape rather than just expanding vertically.
It is instructive to look at roll and pitch angles of the orbits, $\ia$ and $\ib$, which are related to the Kepler angles as follows (valid for prograde orbits, see \citet{Madigan2018}):
\begin{align}
    \ia &= \arcsin{\left(\sin{i} \cos{\omega}\right)}, \\
    \ib &= \arcsin{\left(-\sin{i} \sin{\omega}\right)}, \\
      i &= \arccos\left[\cos \ia \cos \ib \left[1 - (\tan \ia       \tan \ib)^2\right]^{1/2}\right].
\end{align}
The instability is characterized by the exponential growth of the mean $\ia$ and $\ib$ in opposite directions.
We define the timescale of the instability, $t_{\rm insta}$, in our $N$-body simulations as the inverse of the exponential growth rate of the mean $\ib$ of the disk given in units of 
the period of the innermost orbit, $P_i = 2\pi\mu^{-\nicefrac{1}{2}}a_i^{\nicefrac{3}{2}}$, where $\mu = GM_\odot$.
The growth rate is given by the slope of the mean $\ib$ in the linear regime, and it is obtained from with a simple linear, least-squares fit to the log of the mean $\ib$. 
An example is shown in Figure~\ref{fig:growth-rate-compare} where we show the mean $\ib$ for three disks with different $\alpha$.
The linear stage lasts from $\sim\!0.2^\circ$ to $\sim\!10^\circ$, and  instability timescale clearly varies with $\alpha$.
When the instability saturates, mean $i_b$ in the disks is $\sim 20^\circ$ while the mean inclination has reached $35^\circ - 50^\circ$.
Coinciding with the buckling, the mean eccentricity of the disk drops to conserve vector angular momentum of the disk. 
This drop is most pronounced at inner edge of the disk with the mean eccentricity dropping to $0.4 - 0.6$ in this region.
This drop at approximately fixed semi-major axes results in an increase in mean perihelion, $q=a(1-e)$.

\section{Results}

\begin{figure}[!t]
    \centering
    \includegraphics[width=0.95\columnwidth]{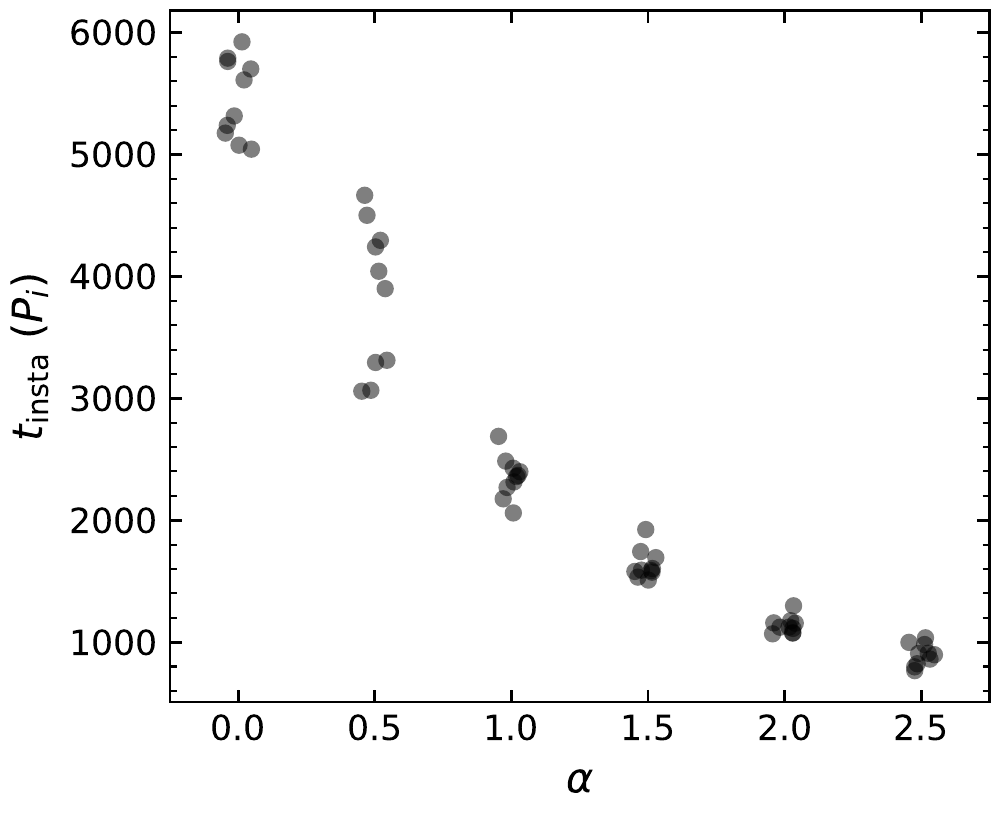}
    \caption{Timescale of the inclination instability, $t_{\rm insta}$, as a function of semi-major axis distribution index, $\alpha$. $N = 400$ and $\Md = 10^{-3}\,M_\oplus$ for all simulations. The instability timescale monotonically decreases with increasing $\alpha$. Jitter is added to reveal overlapping points.}
    \label{fig:timescale-alpha}
\end{figure}

We run 10 $N$-body scattered disk simulations for each $\alpha \in [0, 0.5, 1, 1.5, 2, 2.5]$ and calculate the timescale of the instability as described in Section~\ref{sec:simulations}.
We show the measured timescales as a function of $\alpha$ in Figure~\ref{fig:timescale-alpha}, where we've added jitter to the data to reveal overlapping points.
The instability timescale monotonically decreases with increasing $\alpha$ (steeper density profiles). There is a factor of $\gtrsim \! 6$ difference in timescales across the range in $\alpha$.

Being a secular (orbit-averaged) phenomenon, the inclination instability scales with the secular timescale,
\begin{equation}
    \label{eq:sec-time}
    t_{\rm sec} \sim \frac{M_\odot}{\Md}\,\frac{P}{2\pi}.
\end{equation}
Here we define
$t_{\rm sec}$ as the time it takes to change an orbit's angular momentum by order of its circular angular momentum using a specific torque over one orbital period of $\tau \sim \frac{G M_\odot}{a}$.
In practice, we find that the instability timescale also depends on $N$ due artificially strong two-body scattering present in our low $N$ simulations \citep{Madigan2018}. 
In the limit of infinite $N$, the instability timescale would decrease by a factor of $\sim3$.
The instability timescales  shown in Figure~\ref{fig:timescale-alpha} are for disks with identical $N$ and $\Md$, but different orbital period distributions, $f(P) \propto P^{-\frac{2\alpha + 1}{3}}$. 
The variation of instability timescale with $\alpha$ is due to the variation of $t_{\rm sec}$ with $P$.

\begin{figure}[!t]
    \centering
    \includegraphics[width=\columnwidth]{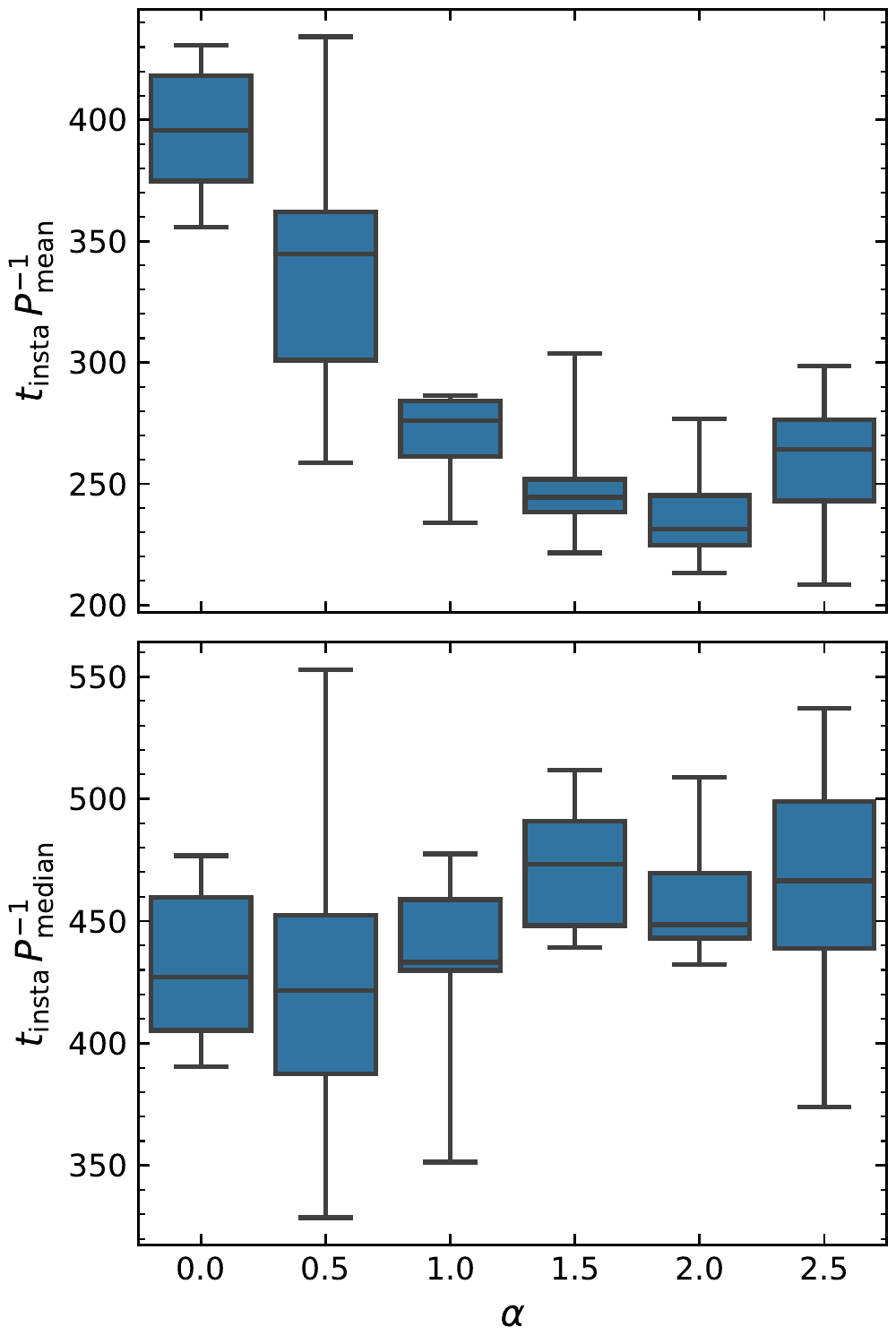}
    \caption{Box and whisker plot of the instability timescale normalized by the mean (top) and median (bottom) period within the disk. The median period properly de-trends the data suggesting the that growth rate at fixed $\Md$ and $N$ is determined by the median period of an orbit in the disk.}
    \label{fig:timescale-normed}
\end{figure}

In Figure~\ref{fig:timescale-normed}, we  normalize the instability timescales shown in Figure~\ref{fig:timescale-alpha} with the mean (top) and median (bottom) orbital periods of the disks.
We use box and whisker plots to summarize the distribution of timescales for each $\alpha$.
The median period successfully de-trends the data to within the inter-quartile range, suggesting that,
\begin{equation}
    t_{\rm insta} \propto t_{\rm sec} \propto P_{\rm median},
\end{equation}
where,
\begin{align}
    P_{\rm median} = 2\pi\mu^{-\nicefrac{1}{2}}
        \begin{cases}
            \left(\frac{1}{2} \left( a_o^{1-\alpha} + a_i^{1-\alpha} \right) \right)^{\frac{3}{2(1-\alpha)}} & \alpha \neq 1\\
            (a_o a_i)^{\nicefrac{3}{4}} & \alpha = 1 \\
        \end{cases},
\end{align}
for $dN \propto a^{-\alpha} da$ in the range $[a_i,a_o]$. 

Previously, we have used $P_i$ to define $t_{\rm sec}$ for our disks, but these results suggest that $P_{\rm median}$ is the correct choice. 
At $\alpha = 0$, the mean and median periods are equal, and as $\alpha$ increases, the ratio of the mean to the median period increases. 
For the power law distributions explored here, the mean period inherently over-weights the larger period orbits while the median weights each orbit equally.
Because the orbits in our simulations have equal mass, the median period is the mass-averaged period.

The variance in $t_{\rm insta}$ at fixed $\alpha$ decreases with increasing $\alpha$ in Figure~\ref{fig:timescale-alpha}.
This variance is due to sampling noise in the median $P$ from simulation to simulation.
The trend in the variance with $\alpha$ is due to the difference in the slope of the cumulative distribution function (CDF) of $P$ at the median $P$ for disks with different $\alpha$.
We define $F(P)$ as the CDF of the period, $\delta P_{\rm median} = P_{\rm median} - P_{S,{\rm median}}$ as the difference between the median period of the underlying distribution and the sample distribution, and $\delta F(P_{\rm median}) = F(P_{\rm median}) - F_S(P_{\rm median})$ as the difference between the underlying CDF and sample CDF at the median of the underlying distribution. We can then approximate,
\begin{equation}
    \delta P_{\rm median} \approx \frac{dP}{dF(P_{\rm median})} \, \delta F(P_{\rm median}).
\end{equation}
$\delta F(P_{\rm median})$ is primarily a function of $N$ which is the same for all our models.
However, $\frac{dP}{dF(P_{\rm median})}$ (the inverse of the slope of the CDF)  decreases with increasing $\alpha$, explaining why the variance in the median $P$ and therefore $t_{\rm insta}$ decreases with increasing $\alpha$.
\section{Discussion}
\label{sec:discussion}

Differential apsidal precession induced  by the giant planets in the trans-Neptunian region suppresses the instability as the coherence time over which orbits can torque each other is reduced.
In \citet{Zderic2020b}, we found that the inclination instability is suppressed when the timescale for the instability and the differential precession timescale are approximately equal. 
We define the timescale ratio as,
\begin{equation}
    r = \frac{t_{\rm insta}}{t_{\rm prec}}.
\end{equation}
We define the differential precession timescale,
\begin{equation}
    t_{\rm prec} = \frac{1\, {\rm rad}}{\dot{\varpi}_{\rm uqrt} - \dot{\varpi}_{\rm lqrt}},
\end{equation}
where $\varpi = \omega + \Omega$ is the longitude of periapsis and $\dot{\varpi}_{\rm uqrt}$ and $\dot{\varpi}_{\rm lqrt}$ are the upper and lower quartile precession rates for the disk from our simulations. This definition emphasizes the importance of differential precession within the disk in weakening the mutual secular torques between the disk orbits, as opposed to the absolute value of the precession rate at any point in the disk \citep{Zderic2020b}.

For a scattered disk orbital distribution and the current configuration of giant planets, we found that the inclination instability was suppressed when $r \gtrsim 1.2 \pm 0.4$.
To obtain $t_{\rm prec}$ for a realistic solar system scattered disk, we scale the measured precession rates in our simulations to lower disk mass using the secular timescale, $\dot{\varpi}_{\rm disk} \propto \Md$.
As our disk is composed of highly eccentric orbits, we cannot use a disturbing function approach to get a close-formed expression for the disk precession rate like those found in \citet{Silsbee2015, Sefilian2021}, and we must calculate the precession timescale numerically.
In addition, we inject apsidal precession modeling the influences of the giant planets in the trans-Neptunian region with a quadrupole-level multipole expansion, i.e.\ $J_2$ precession.
For the inclination instability timescale, we use the $N\rightarrow\infty$ expression for scattered disk found in \citet{Zderic2020b} except now including a $P_{\rm median}$ term normalized to $P_{\rm median}$ when $\alpha=1$,
\begin{equation}
    t_{\rm insta} = \frac{2.4}{\pi} \frac{M_\odot}{\Md} \frac{P_{\rm median}(\alpha)}{P_{\rm median}(1)} P_i.
\end{equation}

\begin{figure}[!t]
    \centering
    \includegraphics[width=\columnwidth]{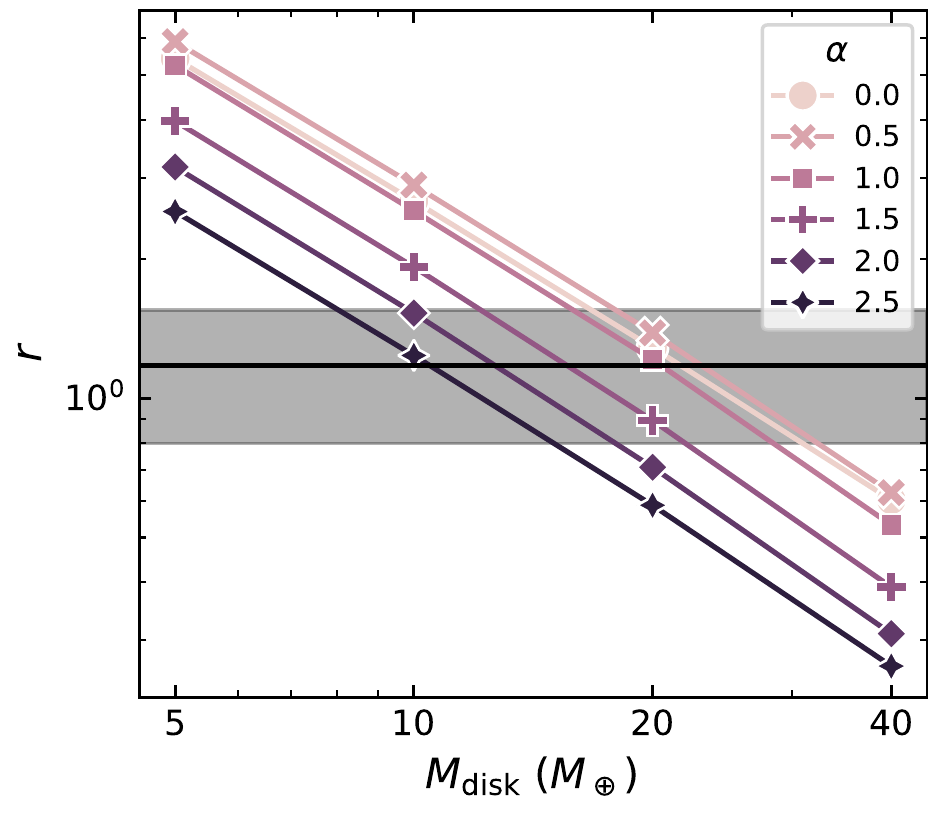}
    \caption{Instability and precession timescale ratio, $r$, as a function of disk mass, $\Md$, calculated from our simulations. $\alpha$ is distinguished by color and marker. The critical $r$ value determining the transition from stability (above) to instability (below) is shown with a grey band. We find that $\alpha \leq 1$ have similar disk mass requirements needing more than $20\,M_\oplus$ to be unstable, while $\alpha > 1$ require less mass to be unstable.}
    \label{fig:r-crit}
\end{figure}

Using these timescales, we calculate $r$ as a function of $\Md$ for different $\alpha$ (see Figure~\ref{fig:r-crit}). 
We find that the dramatic change in growth rate for different disk profiles shown in Figure~\ref{fig:timescale-alpha} is tempered by the difference in differential precession in these disks. 
In fact, we find that the dependence of $r$ on $\Md$ is nearly identical for $\alpha \leq 1$.
At the inner edge of these disks, orbital precession is driven by the giant planets, $\dot{\varpi}_{J_2} \propto a^{-7/2}$. 
The steeper disk profiles place more bodies near the inner edge of the disk, and a larger $t_{\rm prec}$ results.
However, we do find that $r$ decreases with $\alpha$ if $\alpha > 1$.
The differential precession rate in these steeper disks is similar to the $\alpha=1$ disk while $t_{\rm insta}$ still decreases $\propto P_{\rm median}$.
Therefore, the steepest disk profiles need slightly less mass to resist the $J_2$ precession of the giant planets.
The $\alpha = 2.5$ disk profile found by \citet{Huang2022} requires $\Md \gtrsim 10\,M_\oplus$ to be unstable, a factor of two lower than the $\alpha=1$ disks used in our prior works.
This buckled population will depopulate over time as the bodies continue to torque each others' orbits. When torqued to low orbital angular momenta, disk bodies are vulnerable to scattering and removal by giant planets. The rate at which the population decreases is left for future work.

The semi-major axis distribution of the primordial scattered disk effects the post-instability orbital distributions.
In comparison to the $\alpha=1$ distribution explored in our previous work, we find that:
\begin{enumerate} 
    \item shallow (small $\alpha$) disks produce large perihelion orbits post-instability ($q \lesssim 300 \,{\rm au}$) in a broad distribution while steeper disks (large $\alpha$) produce more modestly detached orbits ($q \lesssim 100\,{\rm au}$) in a narrower distribution,
    \item post-instability, mean perihelion distance increases with increasing semi-major axis in smaller $\alpha$ disks while the opposite occurs in larger $\alpha$ disks,
    \item the post-instability inclinations in the smaller $\alpha$ disks are larger ($\sim50 \pm 40\,{\rm deg}$) compared to larger $\alpha$ disks ($\sim35 \pm 20\,{\rm deg}$),
    \item all the unstable disks torque orbits to retrograde orientations, but the small $\alpha$ disks produce them at a higher rate and are capable of producing retrograde orbits before the disk buckles (preliminary results suggest $\sim\!4$ times more retrograde orbits are produced in the $\alpha=0$ disks compared to the $\alpha=2.5$ disks).
\end{enumerate}

We've shown here that the median orbit within the disk determines the instability's timescale. The median orbit in a steeper disk also has a lower initial eccentricity than one in a shallower disk. This means that steeper disks buckle more weakly and result in lower post-instability inclination and perihelion distributions.  
Upcoming observational surveys, such as the Legacy Survey of Space and Time (LSST) on the Vera Rubin Observatory \citep{Ivezic2019}, may be able to distinguish between a steep or shallow primordial scattered disk in the outer solar system using these trends. 
The discovery of detached objects with $q\gtrsim100\,{\rm au}$ at $a\gtrsim500\,{\rm au}$ would rule out unstable primordial scattered disks with $\alpha \gtrsim 2$, and the discovery of extremely detached orbits with $q \sim 300\,{\rm au}$ would require $\alpha \sim 0$.
An abundance of highly inclined and retrograde orbits would indicate a flatter primordial scattered disk semi-major axis distribution.

\acknowledgments
\section*{Acknowledgements}

We thank the referee for their useful comments.
AM gratefully acknowledges support from the David and Lucile Packard Foundation. 
This work utilized resources from the University of Colorado Boulder Research Computing Group, which is supported by the National Science Foundation (awards ACI-1532235 and ACI-1532236), the University of Colorado Boulder, and Colorado State University. 

\software{\texttt{REBOUND} \citep{Rein2012}}

\footnotesize{
\bibliographystyle{aastex}
\bibliography{main.bib}
}

\end{document}